\documentclass[twoside,11pt]{article}

\usepackage{graphicx}

\usepackage{amssymb}
\usepackage{amsmath}
\usepackage{amsfonts}
\usepackage{wasysym}

\usepackage{titlesec}

\usepackage{cite}

\titleformat*{\section}{\bfseries}
\titleformat*{\subsection}{\bfseries}

\begin{document}           

\title{\Large 
The matter-energy intensity distribution in a quantum gravitational system}
\author{V.E. Kuzmichev, V.V. Kuzmichev\\[0.5cm]
\itshape Bogolyubov Institute for Theoretical Physics,\\
\itshape National Academy of Sciences of Ukraine, Kiev, 03680 Ukraine}

\date{}

\maketitle

\begin{abstract}
In the framework of the method of constraint system quantization, a quantum gravitational system (QGS) with
the maximally symmetric geometry is studied. The state vector of the QGS satisfies 
the set of wave equations which describes the time evolution of a quantum system in the space of quantum fields.
It is shown that this state vector can be normalized to unity. The generalization of the wave equations
to the domain of negative values of the cosmic scale factor is made. For the arrow of time from past to future, 
the state vector describes the QGS contracting for the negative values of the scale factor and expanding for its positive values.
The intensity distributions of matter are calculated for two exactly solvable models of spatially closed and flat QGSs formed by
dust and radiation. The analogies with the motion in time of minimum wave packet for spatially closed QGS and with the phenomenon 
of diffraction in optics for flat QGS are drawn.
\end{abstract}


\section{Problem definition and basic equations}\label{sec:1}
The study of matter-energy density distribution in a quantum gravitational system (QGS) is of interest in connection with the
problem of the mechanism of nucleation of the expanding universe from the initial cosmological singularity point, as it is asserted by
the standard cosmological model. This problem, as well as the question about the prehistory of the universe before its nucleation,
can find the solution in quantum theory which considers matter and gravitation as quantum fields. 

The method of constraint system quantization \cite{Dir58,ADM,DeW,Whe,Kuch91,Kuch92,Tor92,Ish,Bro,Kuz02,Kuz08,Kuz08a} 
can be taken as a basis of quantum theory of gravity suitable for the investigation of cosmological and other quantum gravitational systems 
\cite{Lun73,Dem80,Kuz08b,Kuz09,Kuz10,Kuz13,Kuz15,Kuz16}. In this theory,
the state vector (wave function) satisfies the set of wave equations which describes the time evolution of a quantum system in 
a generalized space of quantum fields. The probabilistic interpretation of the state vector of QGS implies its normalizability. 

In the simplest case of the maximally symmetric geometry with the Robertson-Walker metric, the geometric properties of the system
are determined by a single variable, namely the cosmic scale factor $a$.  We will consider the homogeneous isotropic QGS formed by 
matter in the form of a uniform scalar field $\phi$. This field can be interpreted as a surrogate of all possible real physical fields 
of matter averaged with respect to spin, space and other degrees of freedom. 
In addition, it will be accepted that QGS is filled with a perfect fluid in the form of a relativistic matter (further referred as radiation) which 
defines a material reference frame enabling us to introduce the time variable (recognize the instants of time) \cite{Kuch91,Bro,Kuz08}.

It is convenient to formulate quantum theory in terms of dimensionless variables and parameters, in which length is measured in modified units of
Planck length $l_{P} = \sqrt{2 G \hbar / (3 \pi c^{3})}$, proper time is expressed in Planck time units $t_{P} = l_{P} / c$, and mass-energy 
is taken in Planck mass units $m_{P} = \hbar / (l_{P} c)$. The Planck density $\rho_{P} = 3 c^{4} / (8 \pi G l_{P}^{2})$ is used 
as a unit of energy density and pressure. The scalar field is taken in $\phi_{P} = \sqrt{3 c^{4} / (8 \pi G)}$. Here $G$ is Newton's 
gravitational constant. Then the basic equations of the QGS model can be reduced to the following simple set of two differential
equations in partial derivatives for the state vector $\Psi$ \cite{Kuz08,Kuz08a,Kuz13,Kuz15,Kuz16}
\begin{equation}\label{1}
\left(-i \partial_{T} - \frac{2}{3} E \right)\Psi = 0,
\end{equation}
\begin{equation}\label{2}
\left(- \partial_{a}^{2} + \mathrm{k} a^{2} - 2 a H_{\phi} - E \right)\Psi = 0,
\end{equation}
where $T$ is a conformal time expressed in radians. In general relativity, the cosmic scale factor $a$ describes
the overall expansion or contraction of the cosmological system, being the function of proper time $\tau$ which is connected 
with the conformal time $T$ by the equation
\begin{equation}\label{3}
d \tau = a\, dT.
\end{equation}
In Eqs.~(\ref{1}) and (\ref{2}), the quantities $a$, $\phi$ and $T$ are independent variables of the state vector $\Psi = \Psi (a, \phi, T)$.
The parameter $E$ is a real constant which is determined by the energy density of a perfect fluid $\rho_{\gamma}$ taken in the form
\begin{equation}\label{4}
\rho_{\gamma} = \frac{E}{a^{4}}.
\end{equation}
In natural physical units, $E$ has the dimension of [Energy $\times$ Length], $[E] = [\hbar c]$. The coefficient $2/3$ in Eq.~(\ref{1})
is caused by the choice of the parameter $T$ as the conformal time variable.

The operator $H_{\phi}$ in Eq.~(\ref{2}) is the Hamiltonian of the field $\phi$. This Hamiltonian is defined in a curved space-time and
therefore it depends on a scale factor $a$ as a parameter, $H_{\phi} = H_{\phi} (a)$. If the potential term of the uniform scalar field
$\phi$ is described by the scalar function $V(\phi)$, then
\begin{equation}\label{5}
H_{\phi} = \frac{1}{2} a^{3} \rho_{\phi}, \qquad \rho_{\phi} = \frac{2}{a^{6}} \partial_{\phi}^{2} + V(\phi),
\end{equation}
and
\begin{equation}\label{6}
L_{\phi} = \frac{1}{2} a^{3} p_{\phi}, \qquad p_{\phi} = \frac{2}{a^{6}} \partial_{\phi}^{2} - V(\phi),
\end{equation}
where $\rho_{\phi}$ and $p_{\phi}$ are the operators of the energy density and pressure. The $L_{\phi}$ can be interpreted
as the Lagrangian of the scalar field. The variable $\phi$ is defined on the interval: $\phi \in (-\infty,+\infty)$.

In Eq.~(\ref{2}), we single out the curvature constant $\mathrm{k}$, $\mathrm{k} = +1,0,-1$ for for spatially closed, flat, and open QGS.
The derivation of Eqs.~(\ref{1}) and (\ref{2}) \cite{Kuz08,Kuz13} does not depend on the numerical value of $\mathrm{k}$.

The variables $a$ and $\phi$ satisfy the commutation relations
\begin{equation*}
[a,-i \partial_{a}] = i, \qquad [\phi,-i \partial_{\phi}] = i.
\end{equation*}
All other commutators vanish.

The equations (\ref{1}) and (\ref{2}) can be rewritten as one time equation of Schr\"{o}dinger type in the space of two variables
$a$ and $\phi$ with the time-independent operator with the dimension of [Energy $\times$ Length] instead of a Hamiltonian.

\section{Barotropic fluid}\label{sec:2}
It is convenient to pass from $(a, \phi)$-representation of the wave function $\Psi$ to a representation in which a continuous
variable $\phi$ is replaced by a discrete or continuous set of values of quantum number $k$ which characterizes the states 
of the matter field in a comoving volume $\frac{1}{2} a^{3}$. With that end in view, we introduce the complete set of
orthonormalized state vectors of the scalar field $\langle \chi | u_{k} \rangle$ in a representation of a rescaled variable
$\chi = \chi (a, \phi)$, in which the Hamiltonian $H_{\phi}$ is diagonalized,
\begin{equation}\label{7}
\langle u_{k} | H_{\phi} | u_{k'} \rangle = M_{k}(a)\,\delta_{k k'}.
\end{equation}
After averaging of $H_{\phi}$ with respect to the field $\chi$, we transform from the scalar field to the new effective matter
in discrete and/or continuous $k$th state with the proper mass-energy $M_{k}(a) = \frac{1}{2} a^{3} \rho_{m}$.
The energy density $\rho_{m}$ and pressure $p_{m}$ of such an averaged matter are
\begin{equation}\label{8}
\rho_{m} = \langle u_{k} | \rho_{\phi} | u_{k} \rangle, \qquad p_{m} = \langle u_{k} | p_{\phi} | u_{k} \rangle.
\end{equation}
Introducing the equation of state parameter
\begin{equation}\label{9}
w_{m}(a) = - \frac{1}{3} \frac{d \ln M_{k} (a)}{d \ln a},
\end{equation}
we find that the averaged matter is a barotropic fluid with the equation of state 
\begin{equation}\label{10}
p_{m} = w_{m}(a) \rho_{m}.
\end{equation}

The explicit form of the field $\chi$ and the mass-energy $M_{k} (a)$ are different for the different functions $V(\phi)$.
If, for example, $V(\phi) = \lambda_{\alpha} \phi^{\alpha}$, where $\lambda_{\alpha}$ is the 
coupling constant, $\alpha$ takes arbitrary non-negative values, and the summation with respect to $\alpha$ is not assumed,
then \cite{Kuz13} 
\begin{equation}\label{11}
\chi = \left(\sqrt{2 \lambda_{\alpha}} \frac{a^{3}}{2} \right)^{\frac{2}{2 + \alpha}} \phi, \quad
M_{k}(a) = \epsilon_{k} \left(\frac{\lambda_{\alpha}}{2} \right)^{\frac{2}{2 + \alpha}} a^{\frac{3 (2 - \alpha)}{2 + \alpha}}, \quad
w_{m}(a) = \frac{\alpha - 2}{\alpha + 2},
\end{equation}
where $\epsilon_{k}$ is an eigenvalue of the equation
\begin{equation}\label{12}
(-\partial_{\chi}^{2} + \chi^{\alpha} - \epsilon_{k}) |u_{k} \rangle = 0.
\end{equation}

In a particular case of the $\phi^{2}$-model, the parameter $w_{m}(a) = 0$ and the barotropic fluid becomes an aggregate of 
separate macroscopic bodies (dust) with the mass-energy $M_{k} = \sqrt{2 \lambda_{2}}(k + \frac{1}{2})$, which does not depend on $a$,
and $\epsilon_{k} = 2k +1$, where $k=0,1,2, \ldots$. The equation (\ref{12}) for $\alpha = 2$ is the equation for quantum oscillator.
The mass $M_{k}$ can be interpreted as a sum of masses of separate excitation quanta of the spatially coherent oscillations
of the field $\chi$ about the equilibrium state $\chi = 0$. The quantum number $k$ is the number of these excitation quanta
with the mass $\sqrt{2 \lambda_{2}}$. Taking the mass of proton $ \approx 1$ GeV ($\sim 10^{-19}$ in Planck mass units) as such a mass,
one obtains $M_{k} \sim 10^{80}$ GeV ($\sim 10^{-61}$) when the number of protons $k  \sim 10^{80}$. Such a mass of a dust leads to
the actual density of matter in the observed part of our universe \cite{PDG}. The review of the properties of 
the barotropic fluid in the  $\phi^{\alpha}$-models with $\alpha \neq 2$ is given in Refs. \cite{Kuz13}.

\section{General solution}\label{sec:3}
Using the completeness and orthonormality of the state vectors $\langle \chi | u_{k} \rangle$, the state vector $\Psi$ of QGS
in $(a, \chi)$-representation can be represented in the form of the superposition of all possible $k$th states of the barotropic fluid
\begin{equation}\label{13}
\Psi = \sum_{k} | u_{k} \rangle \langle u_{k}| \Psi \rangle,
\end{equation}
where $\langle u_{k}| \Psi \rangle \equiv \psi_{k}(a, T)$ satisfies the differential equations
\begin{equation}\label{14}
\left(-i \partial_{T} - \frac{2}{3} E \right)\psi_{k}(a, T)= 0,
\end{equation}
\begin{equation}\label{15}
\left(- \partial_{a}^{2} + \mathrm{k} a^{2} - 2 a M_{k}(a) - E \right)\psi_{k}(a, T) = 0.
\end{equation}
The general solution of this set has the form
\begin{equation}\label{16}
\psi_{k}(a, T) = \sum_{n} c_{n k} (T) f_{n k}(a) + \int dE\, c_{E k} (T) f_{E k}(a),
\end{equation}
where
\begin{equation}\label{17}
c_{n k} (T) = c_{n k} (T_{0}) \exp \left\{i \frac{2}{3} E_{n} (T - T_{0}) \right\},
\end{equation}
\begin{equation}\label{18}
c_{E k} (T) = c_{E k} (T_{0}) \exp \left\{i \frac{2}{3} E (T - T_{0}) \right\}.
\end{equation}
The wave functions $f(a)$ in Eq.~(\ref{16}) satisfy the equation
\begin{equation}\label{19}
\left(- \partial_{a}^{2} + \mathrm{k} a^{2} - 2 a M_{k}(a) - E \right) f(a) = 0,
\end{equation}
where $f(a) = f_{n k}(a)$ and $E = E_{n}$ for a discrete $n$th state of radiation with the density (\ref{4}), and
 $f(a) = f_{E k}(a)$ for a continuous $E$th state of radiation. In both cases, the QGS is considered as a material point
 moving in the potential function
\begin{equation}\label{20}
U_{k}(a) = \mathrm{k} a^{2} - 2 a M_{k}(a)
\end{equation}
formed by the barotropic fluid in $k$th state.

The parameter $T_{0}$ in Eqs.~(\ref{17}) and (\ref{18}) is an arbitrary constant taken as a time reference point. The equation (\ref{19})
determines the stationary quantum state of QGS at some fixed instant of time $T_{0}$, the choice of which is arbitrary, 
$f(a) \equiv f(a, T_{0})$.

In probabilistic interpretation of quantum theory, the coefficient $c_{n k} (T_{0})$ gives the probability $|c_{n k} (T_{0})|^{2}$ to find
the QGS in $n$th state of the discrete spectrum of radiation and $k$th state of the barotropic fluid at the instant of time $T_{0}$.
The same interpretation can be given for the coefficient $c_{E k} (T_{0})$ in the case of continuous state of radiation.

The conditions of normalization and orthogonality can be imposed on the wave functions $f(a)$,
\begin{equation}\label{21}
\langle f_{n k} | f_{n',k} \rangle = \delta_{n n'}, \quad \langle f_{E k} | f_{E',k} \rangle = \delta (E - E'), \quad \langle f_{n k} | f_{E,k} \rangle = 0.
\end{equation}
Then the state vector $\Psi$ appears be normalized to unity,
\begin{equation}\label{22}
\langle \Psi | \Psi \rangle = \sum_{k} \left(\sum_{n} |c_{n k} (T_{0})|^{2} + \int dE\, |c_{E k} (T_{0})|^{2} \right) = 1,
\end{equation}
under the condition that the probability summed over all possible quantum states of radiation and barotropic fluid equals to unity,
i.e. the QGS with the state vector $\Psi$ exists.

\section{Generalization}\label{sec:4}
According to Eq.~(\ref{5}), the equation (\ref{2}) is invariant under the inversion $a \rightarrow -a$. Taking into account that
the Robertson-Walker line element contains only even powers of $a$ and also that the sign of $a$ has not been fixed when deriving 
Eqs.~(\ref{1}) and (\ref{2}), the equation (\ref{19}) can be generalized by extending to the domain of negative values of $a$,
so that $a \in (-\infty,+\infty)$ (cf. Refs.~\cite{Kuz97,Kan}).

\begin{figure}[ht]
\begin{center}
\includegraphics*[width=0.7\textwidth]{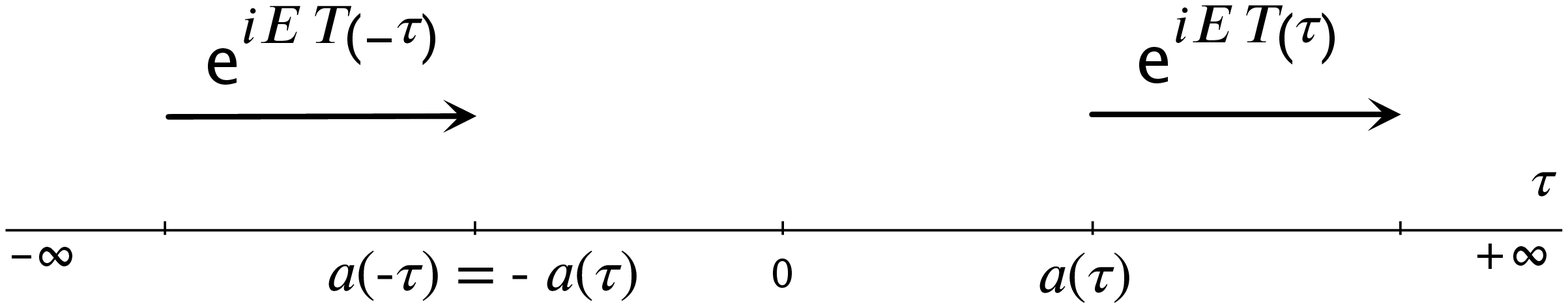}
\end{center}
\caption{The wave expanding in proper time $\tau$.} \label{fig:1}
\end{figure}

In order to clarify the physical meaning of the solutions of Eq.~(\ref{19}) in the domain $a < 0$, let us integrate Eq.~(\ref{3}),
\begin{equation}\label{23}
T(\tau) = T_{0} + \int_{0}^{\tau} \frac{d \tau'}{a(\tau')}, \quad \mbox{at}\ T(0) = T_{0}.
\end{equation}
It gives
\begin{equation}\label{24}
T(\tau) = T(-\tau) , \quad \mbox{at}\ a(-\tau) = -a(\tau).
\end{equation}
From Eqs.~(\ref{13})-(\ref{18}), it follows that the dependence of the state vector $\Psi$ on time $T(\tau)$ is determined by the exponential
multiplier $\exp(i E T)$, where the inessential multiplier $2/3$ is omitted and the natural condition $T_{0} = 0$ is imposed. Let this exponential
describes the wave expanding from $\tau = - \infty$ in the direction $\tau = + \infty$ and passing through the point $\tau = 0$, where
$a(0) = 0$ according to Eq.~(\ref{24}). An illustration is given in Fig.~\ref{fig:1}.

Then the scale factor $a \in (-\infty,0]$ corresponds to the values $\tau \in (-\infty,0]$, and the scale factor $a \in [0, +\infty)$ corresponds to
the values $\tau \in [0, +\infty)$. As a result for the arrow of time from $\tau  = -\infty$ to $\tau  = +\infty$, the state vector $\Psi$
describes the QGS contracting on the semiaxis $a < 0$, since $|a|$ decreases, and the QGS expanding on the semiaxis $a > 0$,
because $|a|$ increases.

The instant of time $\tau = 0$ can be interpreted as the instant of nucleation of the quantum system expanding
in time from the point $a = 0$, although any nucleation ``from nothing'' does not occur physically.
What  happens at the instant $\tau = 0$ is that the regime of the preceding contraction of the system changes into the subsequent
expansion. The equation (\ref{19}) describes the stationary states of QGS for a given value of constant $E$. The state vector (\ref{13})
contains all information about the system as a whole: the cross-section $|a| = const$ determines the quantum state of QGS
at the time instant $\tau$, when such a value of the scale factor holds.

If one applies the above-described scenario to our universe at the Planck epoch, interpreting the passage through the point $a = 0$ at
$\tau = 0$ as the nucleation of expanding universe with $a > 0$ at $\tau > 0$, then the answer to the question ``What was with the 
quantum system before the instant of nucleation of the universe of our (expanding) type?'' can be given: there has existed another
universe with the same mass-energy $M_{k}(a)$ and wave function $f(a)$ characterized by the same quantum numbers for matter and radiation
as the nucleated universe, however that universe has been contracting up to the state with $a = 0$, which not necessarily will be
singular (see Sect.~5 below).

\section{Intensity distribution}\label{sec:5}
Let us consider QGS in which the barotropic fluid (matter) and radiation are in some definite quantum states. In such a quantum system,
the intensity distribution of matter-energy as a function of $a$ is given by the expression
\begin{equation}\label{25}
I(a) = M(a) |f(a)|^{2},
\end{equation}
where indices of the states of matter ($k$) and radiation ($n$) are omitted. The wave function $f(a)$ is the solution of Eq.~(\ref{19})
complemented with the appropriate boundary conditions which determine, for example, the behaviour of $f(a)$ in the
asymptotic domain of large values of $|a|$.

The intensity summed over all possible values of $a$ gives the mean mass-energy of matter in QGS in the state $f(a)$ normalized to unity,
\begin{equation}\label{26}
\langle M(a) \rangle = \int da\, I(a).
\end{equation}
We note that $\langle M(a) \rangle = M = const$ for the $\phi^{2}$-model. As an example, we will calculate the intensity (\ref{25})
in the exactly solvable models of spatially closed and flat QGS filled with dust and radiation.

\subsection{Spatially closed QGS}\label{sec:5.1}
In the model of spatially closed QGS formed by dust whose mass does not depend on $a$, $M(a) = M = const$, the equation (\ref{19})
is reduced to the equation for an oscillator by substitution of the variable $a = \xi + M$,
\begin{equation}\label{27}
\left(- \partial_{\xi}^{2} + \xi^{2} - \epsilon \right) f(\xi) = 0,
\end{equation}
where an eigenvalue $\epsilon = E + M^{2}$. Changing from variable $a$ to $\xi$ restores inversion invariance of Eq.~(\ref{19})
violated in the case when matter is represented by dust. The variable $\xi$ describes the deviation of $a$ from its equilibrium value
at the point $a = M$. This variable lies in the interval $[-M,+\infty)$, if $a$ takes only positive values and zero. For $M \gg 1$,
the interval of change of $\xi$ in the normalization integral for the function $f(\xi)$ can be extended to the whole semiaxis of the
negative values of $\xi$. The error arising here is $\sim O((2 M)^{2n - 1} \exp(-M^{2}))$, where $n = 0,1,2,\ldots$ \cite{Kuz08,Kuz08a}.
Then the solution of Eq.~(\ref{27}), decreasing at $|\xi| = + \infty$ and normalized to unity, gives the intensity distribution of dust matter
in the form
\begin{equation}\label{28}
I_{n}(a) = \frac{M}{2^{n} n! \sqrt{\pi}}\, e^{-(a - M)^{2}} H_{n}^{2}(a - M),
\end{equation}
where $H_{n}(\xi)$ is the Hermite polynomial and the variable $a$ takes any values in the interval $(-\infty,+\infty)$. The constant $E$
is quantized in accordance with the condition
\begin{equation}\label{29}
E = 2 n + 1 - M^{2}.
\end{equation}
It takes a sequence of discrete positive values for $2 n + 1 > M^{2}$ and discrete negative values for $2 n + 1 < M^{2}$.
In the latter case, radiation as a perfect fluid is characterized by the negative energy density (\ref{4}) and pressure 
$p_{\gamma} = - \frac{1}{3} |\rho_{\gamma}|$, i.e. a perfect fluid acquires the properties of the antigravitating matter
at small quantum numbers $n$ and large masses $M$. 

The equation (\ref{28}) determines the intensity distribution of dust matter in QGS both in the regime of its contraction, when
$a < 0$, and in the regime of expansion, when $a > 0$. The quantities $n$ and $M$ in Eqs.~ (\ref{28}) and  (\ref{29}) are
free parameters. 

\begin{figure}[ht]
\begin{center}
\includegraphics*[width=0.7\textwidth]{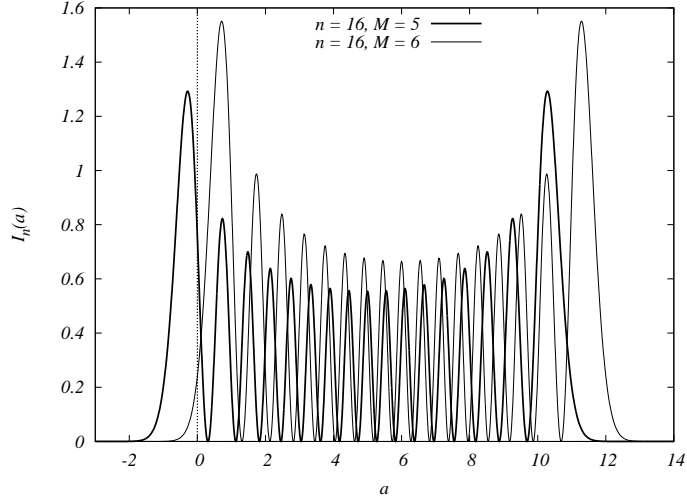}
\end{center}
\caption{The intensity distribution (\ref{28}) for the cases $\rho_{\gamma} > 0$ (bold line) and $\rho_{\gamma} < 0$ (thin line).} \label{fig:2}
\end{figure}

In Fig.~\ref{fig:2}, it is shown the intensity distribution for the parameters $n = 16$ and $M = 5$ (bold line), when $\rho_{\gamma} > 0$,
and $M = 6$ (thin line), when $\rho_{\gamma} < 0$. In the case of the positive energy density of radiation, the intensity evolves so that
the first maximum of $I_{n}(a)$ is reached in the domain of contraction, straight before the boundary point $a = 0$, where the regime of 
contraction changes into expansion. It is as if QGS accumulates the energy just before the beginning of expansion. The intensity at the 
point $a = 0$ is found to be finite ($I_{n}(0) = 0.7$ in Fig.~\ref{fig:2}). In the case $\rho_{\gamma} < 0$, the negative pressure of radiation
pushes out the first maximum into the domain of expansion near the point $a = 0$. In both cases, the intensity oscillates between 
maximum values and zero in the domain of expansion. The value $a = M$ corresponds to the smallest maximum and $I_{n} (0) = I_{n}(2 M)$.
For $a \gg 2 M$, the intensity decreases exponentially.

For $n \gg 1$, the intensity distribution oscillates according to the law
\begin{equation}\label{30}
I_{n}(a) = \frac{M}{\pi} \sqrt{\frac{2}{n}} \cos^{2} \left(\sqrt{2 n} (a - M) - \frac{n \pi}{2} \right).
\end{equation}
Averaging with respect to oscillations gives the intensity which does not depend on $a$,
\begin{equation}\label{31}
\bar{I}_{n}(a) = \frac{M}{\pi} \sqrt{\frac{1}{2 n}} \quad \mbox{at}\ n \rightarrow \infty.
\end{equation}
The general behaviour of the intensity (\ref{28}) reproduces a position probability density of a harmonic oscillator with respect to the
variable $a$ renormalized by the mass $M$. For the QGS under consideration, the equation (\ref{27}) is exact. Its application to the cosmological
problem leads to nontrivial conclusions about the evolution of the intensity distribution of matter in the QGS with all possible parameters 
$n$ and $M$ close to the Planck epoch.

\subsection{Spatially flat QGS}\label{sec:5.2}
\begin{figure}[ht]
\begin{center}
\includegraphics*[width=0.7\textwidth]{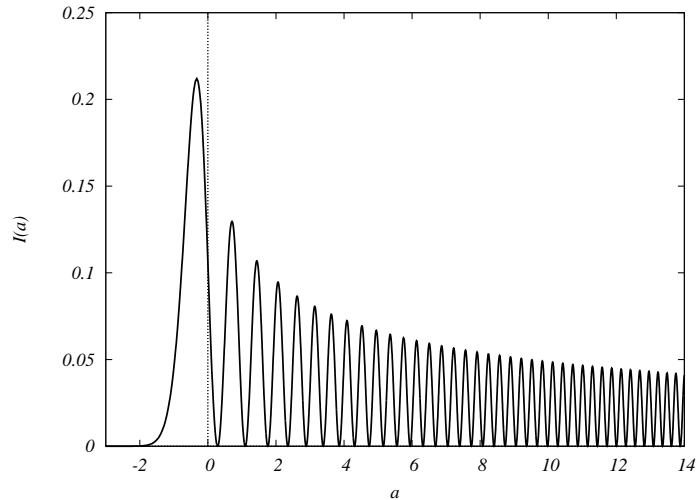}
\end{center}
\caption{The intensity distribution (\ref{34}) for $M = 5$, $E = 8$.} \label{fig:3}
\end{figure}

The prediction about the exponential decreasing of the intensity distribution of matter for $a \gg 2 M$ is made in the model of spatially closed
QGS. Now let us consider another exactly solvable model, namely that of flat space. In the model of spatially flat QGS formed by dust
with the constant mass $M$, the equation (\ref{19}) is reduced to
\begin{equation}\label{32}
\left(\partial_{\xi}^{2} + \xi \right) f(\xi) = 0
\end{equation}
by introducing the variable
\begin{equation}\label{33}
\xi = (2 M)^{1/3} \left(a + \frac{E}{2 M} \right).
\end{equation}
Its general solution is a linear superposition of Airy functions $Ai$ and $Bi$.  We will look for the solution which has a form 
of the outgoing wave at $\xi > 0$ and satisfies the boundary condition $f(-\infty) = 0$. By normalizing this solution to delta-function
as in Eq.~(\ref{21}), we get the following expression for the intensity distribution of matter (\ref{25})
\begin{equation}\label{34}
I(a) = \frac{(2 M)^{2/3}}{2 \pi} Ai^{2}\left(- (2 M)^{1/3} \left(a + \frac{E}{2 M}\right)\right).
\end{equation}
In Fig.~\ref{fig:3}, it is depicted the intensity $I(a)$ (\ref{34}) as a function of $a$ for the same parameters as in Fig.~\ref{fig:2}
for $\rho_{\gamma} > 0$.

As in the case of spatially closed QGS, the intensity (\ref{34}) increases exponentially with contraction of the system, reaching
a maximum, passing through the point $a = 0$ with the finite value ($I(0) = 0.1$) and then oscillating. From the asymptotic 
expression for $Ai(-\xi)$, it follows that the intensity averaged over the oscillations decreases with $a$ according to the law
\begin{equation}\label{35}
\bar{I}(a) = \frac{1}{4 \pi} \sqrt{\frac{2 M}{a}} \quad \mbox{at}\ a \gg \frac{E}{2 M}.
\end{equation}
If one assumes that during the quantum epoch the intensity (\ref{35}) decreases in time as $\tau^{-2}$ like the energy density 
in general relativity, then the scale factor should increase in time according to a power law $a \sim \tau^{4}$. Such a growth of $a$ 
corresponds to an inflationary model, in which it is supposed that the scale factor increases more slowly than in the exponential 
regime \cite{Lyth}.

\section{Analogies}\label{sec:6}
The intensities (\ref{28}) and (\ref{34}) obtained from exact solutions of Eq.~(\ref{19}) allows us to draw analogy 
with known phenomena which are described by the equations of quantum mechanics and optics. We will consider some of these
analogies in detail.

\subsection{Oscillating wave packet}\label{sec:6.1}
The formula (\ref{28}) gives the intensity distribution of matter in stationary states. Let us find how this intensity changes
with proper time in the expanding QGS.

The eigenfunction of the ground state of an oscillator (\ref{27})
\begin{equation}\label{36}
f_{0}(a - \langle a \rangle) = \frac{1}{\pi^{1/4}}\, e^{-\frac{1}{2}(a - \langle a \rangle)^{2}}
\end{equation}
has the form of the normalized minimum packet whose center of gravity is displaced in the positive $a$ direction by an amount
$\langle a \rangle = M$ \cite{Kuz08,Kuz08a}. We assume that such a state corresponds to the time instant 
$\tau = 0$, $\psi (a, \tau = 0) = f_{0}(a - \langle a \rangle)$.
If the equivalent classical system evolves in time $\tau$ with a power-law scale factor, $a = \beta \tau^{\alpha}$, where $\alpha$ and
$\beta$ are some positive constants, then the time phase in Eq.~(\ref{17}) at $T_{0} = 0$ takes the form
\begin{equation}\label{37}
\frac{2}{3} E_{n} T = \left[\frac{1}{2} (1 - M^{2}) + n\right] \omega \tau,
\end{equation}
where the frequency
\begin{equation}\label{38}
\omega = \frac{4}{3 (1 - \alpha) a}
\end{equation}
depends on $a$. From the requirement $\omega > 0$, it follows the restriction: $a > 0$ and $0 \leq \alpha < 1$.
Under these conditions we have $\lim_{\tau \rightarrow 0} \omega \tau = 0$. Taking into account only the sum with respect to $n$
in Eq.~(\ref{16}) (continuos spectrum is absent), using the representation (\ref{17}) with $T_{0} = 0$, and the explicit form of
the function (\ref{36}), as well as the representation of the phase (\ref{37}), one can calculate the wave function
$\psi(a, \tau) = \psi_{k}(a, T(\tau))$. As a result, the intensity distribution $I(a, \tau) = M |\psi(a, \tau)|^{2}$ of matter
with the mass $M$ in wave packet moving in time appears to be equal
\begin{equation}\label{39}
I(a, \tau) = \frac{M}{\sqrt{\pi}} \exp \left\{-\left[a - (1+ \cos \omega \tau) M\right]^{2} \right\}.
\end{equation}
If the mass of matter (dust) goes to zero, then the wave functions tends to the eigenfuction corresponding to the lowest energy of
radiation $\frac{1}{2} \omega$ at $ n = 0$, while the intensity does not depend on time and goes to zero as 
$I(a, \tau) \rightarrow (M/\sqrt{\pi}) \exp(- a^{2})$ at $M \rightarrow 0$. If the mass $M \neq 0$, then the condition $E > 0$
is ensured by including the stationary states with the quantum numbers $n \neq 0$ in the packet. In the case $n \gg 1$,
the main contribution to the packet is made by the eigenfunction with $n = \frac{1}{2} M^{2}$, and the intensity is described
by the expression (\ref{39}).

\begin{figure}[ht]
\begin{center}
\includegraphics*[width=0.9\textwidth]{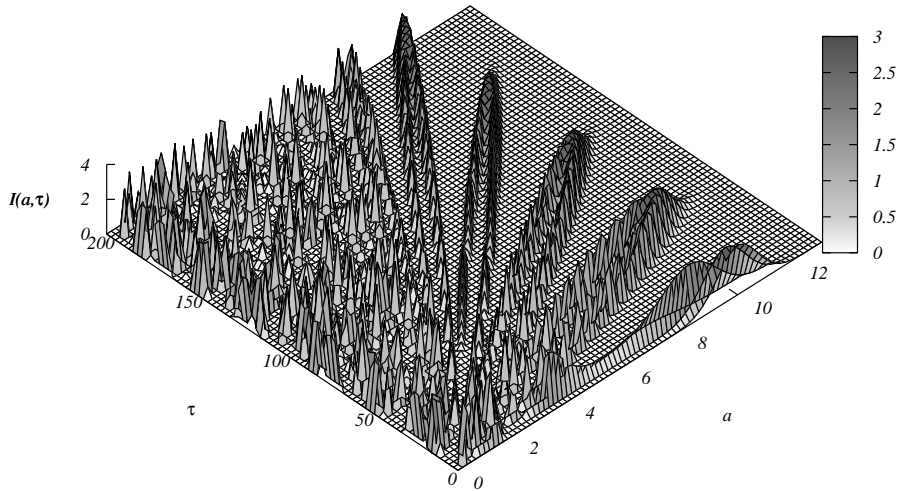}
\end{center}
\caption{The intensity distribution (\ref{39}) for $M = 5$, $\alpha = 0$.} \label{fig:4}
\end{figure}

In Fig.~\ref{fig:4}, it is shown the intensity $I(a, \tau)$ (\ref{39}) for the parameters $M = 5$, and $\alpha = 0$ in Eq.~(\ref{38}).
The mods with $n = 12$ and $n = 13$ make the most important contribution to $I(a, \tau)$. Matter is distributed in $a$ and $\tau$
in the form of periodic structures like petals or stretched bubbles and displaced to their edges. These structures are limited by
the value $a = 2 M$ with respect to $a$ (as in Fig.~\ref{fig:2}), and their number increases with time.

\subsection{Diffraction}\label{sec:6.2}
For the spatially flat QGS formed by dust and filled with radiation, described by Eq.~(\ref{19}) with $\mathrm{k} = 0$ and
$M = const$, an analogy with the motion of a particle in a uniform external field along the coordinate $a$ with the energy $E$
under the action of the force $F = 2 M$ is obvious \cite{Lan3}. Therefore we consider a more interesting analogy with the
distribution of the light intensity in the neighbourhood of the point where its ray is tangent to the caustic \cite{Lan2}.
From Eq.~(\ref{34}), it follows the asymptotic expression for $I(a)$ at $\left(a + \frac{E}{2 M}\right) \gg 1$, which it is
convenient to rewrite in the form
\begin{equation}\label{40}
I \approx \frac{2 A}{\sqrt{-x}} \sin^{2} \left(\frac{2}{3} (- x)^{3/2} \sqrt{\frac{2 \kappa^{2}}{R}} +  \frac{\pi}{4}\right),
\end{equation}
where we denote
\begin{equation}\label{41}
A = \frac{\sqrt{2 M}}{4 \pi}, \quad x = - \left(a + \frac{E}{2 M}\right), \quad \kappa^{2} = \frac{E}{2 a}, \quad R = \frac{E}{2 a M}.
\end{equation}
The introduced quantities have a clear physical meaning. The amplitude $A$ is the intensity far from the caustic which would be obtained
from geometrical optics neglecting diffraction effects, $x$ is the distance from the point of observation along the normal to the caustic
which takes positive values for points on the normal in the direction of its center of curvature, $\kappa^{2}$ is the energy of the ray of light,
$R$ is the radius of curvature of the caustic at the point of observation.

The equation (\ref{40}) describes the intensity for the ray of light at large negative values of $x$. In the radiation-dominated epoch,
$a \ll E/(2 M)$, the radius of curvature is $R \gg 1$. It means that the wave surface is practically flat. In the matter-dominated epoch,
$a \gg E/(2 M)$, we have $R \ll 1$. For $R \sim 0$, the wave surface is practically spherically symmetric, and its center of the caustic 
coincides with the focus. Far from the focus, the averaged intensity decreases as $a^{-1/2}$ with account of diffraction effects (see
Eq.~(\ref{35})). Diffraction in the QGS can be explained by scattering of electromagnetic waves of radiation on massive dust particles
playing the role of opaque bodies (screens). The observed diffraction picture is similar to that depicted in Fig.~\ref{fig:3}.

\end{document}